\documentclass{iau}
\usepackage{graphicx}

\title[JD 11.Pre-solar grains and AGB stars]
{The quest for collapsed/frozen stars in \\ single-line spectroscopic binary systems}

\author[Virginia Trimble]
{Virginia Trimble}

\affiliation{University of California Irvine, \\ 
Department of Physics and Astronomy \\ Irvine, California 92697, USA \\ email: {\tt vtrimble$@$uci.edu} \\[\affilskip]}

\pubyear{2014}
\volume{308} 
\pagerange{119--126}
\jname{The Zel\'dovich Universe \\ Genesis and Growth of the Cosmic Web}
\editors{A.C. Editor, B.D. Editor \& C.E. Editor, eds.}

\begin{document}
\maketitle
\begin{abstract}
Black holes are now commonplace, among the stars, in Galactic centers, and perhaps other places.  But within living memory, their very existence was doubted by many, and few chose to look for them.  Zel\'dovich and Guseinov were first, followed by Trimble and Thorne, using a method that would have identified HDE 226868 as a plausible candidate, if it had been in the 1968 catalogue of spectroscopic binaries.  That it was not arose from an unhappy accident in the observing program of Daniel M. Popper long before the discovery of X-ray binaries and the identification of Cygnus X-1 with that hot, massive star and its collapsed companion.

\end{abstract}

\section{A Contextual Introduction}
The context in which this presentation was assembled and written up includes a recent book called {\it Discovery and Classification in Astronomy: Controversy and Consensus} (Dick 2013) in which the author states at least twice that black holes were not taken seriously by astronomers until the mid 1970's or later.  Now the two widely-cited pioneering papers on powering of quasars by accretion on a central supermassive, compact object came from E.E. Salpeter (1964) and Ya. B. Zel\'dovich \& I.D. Novikov (1964b) which had me doubting the claims immediately.  This was reinforced by the sudden memory that one of my own papers (Trimble \& Thorne 1969) took them seriously rather earlier than the mid-1970's, and that it had its origins in an idea from Zel\'dovich \& Guseinov (1966).  There followed a resolve to look back at this topic and offer an oral version of it to the June, 2014 Zel\'dovich Centenary conference in Moscow.  Complexities associated with the June 15th end of the school year at UCI kept me from attending that meeting, and the organizers of IAU Symposium 308 the next week generously made room for ``The Quest for Collapsed Stars'' just before the conference dinner on the second to last day of the meeting.

\section{Cast of Characters}
A Newtonian version of the story can be pushed back to publications in the late 18th century, when John Michell and Pierre-Simon de Laplace noted the possibility of astronomical entities with escape velocity equaling or exceeding the speed of light.  Relativity enters the picture with a pair of much-cited, but rarely read, papers by Karl Schwarzschild (1916a, b,) that defined what we call the Schwarzschild radius and was already being called the gravitational radius when Oppenheimer and Snyder (1939)  wrote ``On continued gravitational contraction.''

They wrote, and we all now agree, that a sufficiently massive star, having exhausted its thermonuclear sources of energy, must either reduce its mass to of order that of the sun or suffer continuous contraction.  Reaching the gravitational radius takes a day or two for an observer riding the star, but forever as seen from infinity, hence the phrase ``frozen star'' (which came over as ``cooled star'' in some translations).  The last photon that reaches outside observers, however, departs within days, not millennia.

Then there was a war, with very little work on general relativity and its implications being done anyplace for many years.  On the far side of that abyss, John Wheeler and a few Princeton students and colleagues began exploring the territory near very massive compact objects, while in the USSR, two critical early papers predicting inescapable collapse were published by Zel\'dovich \& Novikov (1964, 1965).  We interrupt this tale to explain that those names are in alphabetical order, in Russian, where the letter that English-speakers transcribe as Z looks sort of like a backwards Greek epsilon and comes early in the alphabet, while N looks like H and comes later.

The oral presentation in Tallinn included a number of images of people, publications, and other pieces of paper not readily reproducible here, and also a Sidney Harris cartoon (caption:  It's black, and it looks like a hole.  I'd say it's a black hole) still under copyright.  Figures 1 and 2 are an inscription from the flyleaf of a 1985 book and a 1986 post-New Years letter, and are included to demonstrate that I have a right to tell Zel\'dovich stories.
\\
\begin{figure}
% \vspace*{-2.0 cm}
\begin{center}
\includegraphics{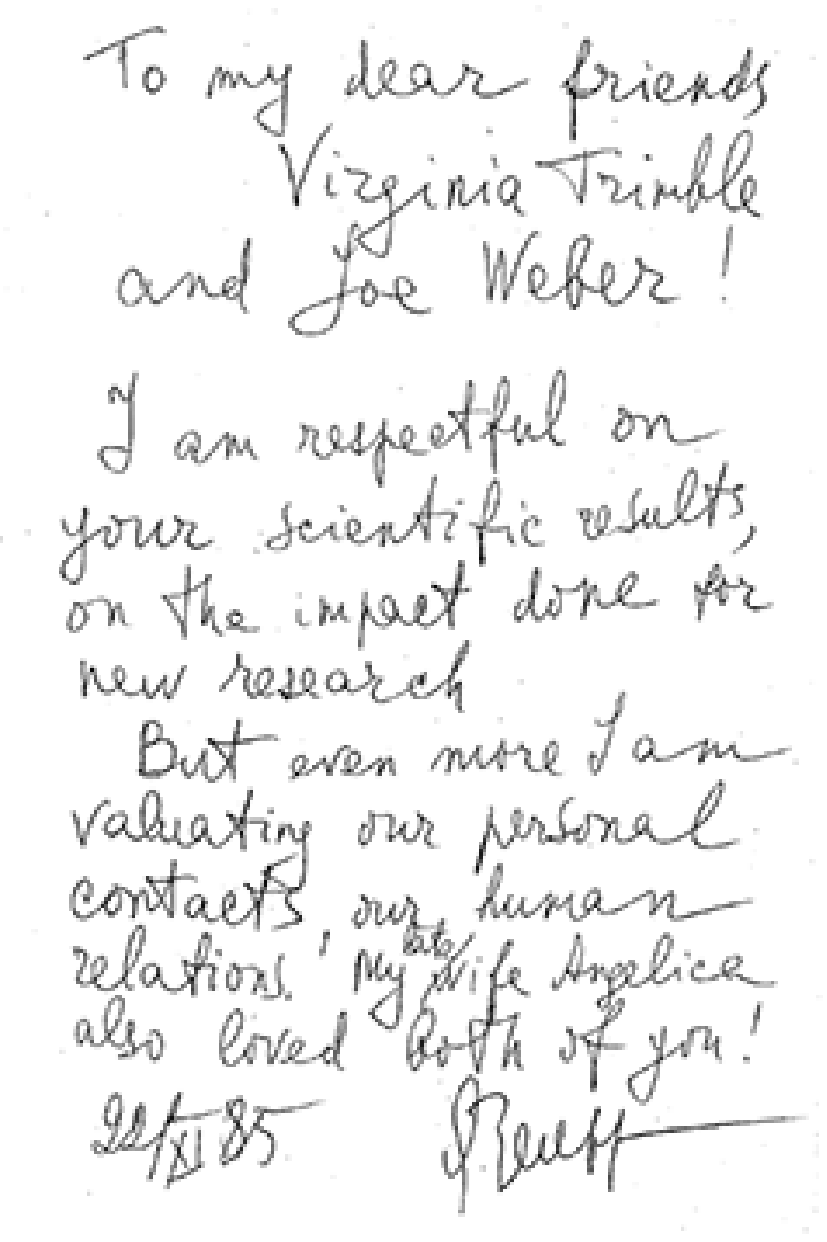} 
% \vspace*{-1.0 cm}
\caption{Fly-leaf inscription on a copy of a 1985 volume of reprints of many of Zel\'dovich's significant papers across many fields, together with a fairly complete list of all his publications up to that time.  Nearly all in Russian, of course.}
\label{fig1}
\end{center}
\end{figure}

\begin{figure}
% \vspace*{-2.0 cm}
\begin{center}
\includegraphics{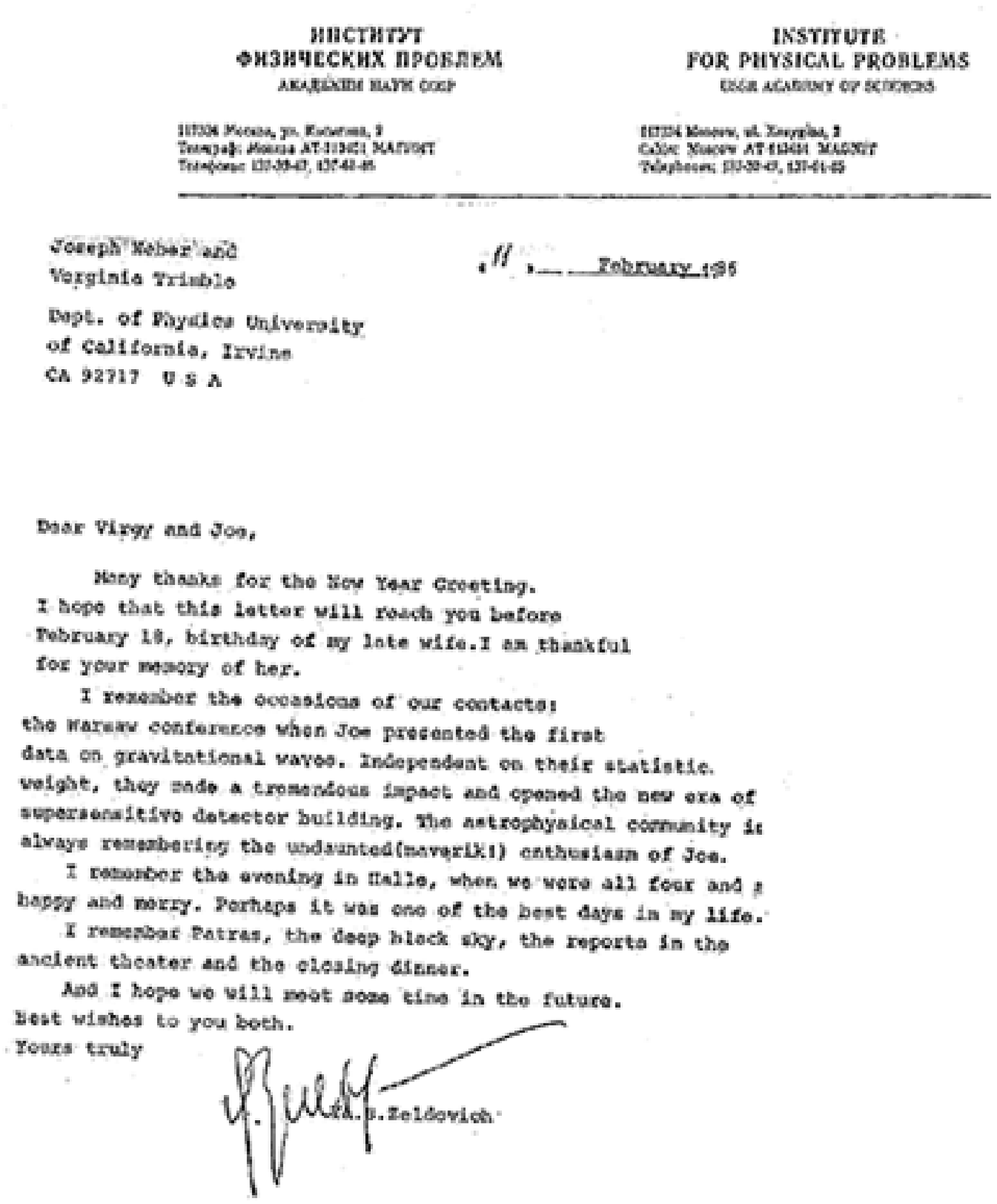} 
% \vspace*{-1.0 cm}
\caption{Copy of a February, 1986 letter recalling the three (and only three) occasions on which he and Joe and/or I had been together.  A somewhat earlier letter told us of Angelica's death and instructed Joe to take good care of ``Vergie.''  A later one expressed regret that we had not been at his NAS talk, and I have always hoped, though never been quite sure, that our response arrived before he died.}
\label{fig2}
\end{center}
\end{figure}
The events referred to near the end of the letter are (1) the GRG meeting in Warsaw in 1963, from which Richard Feynman wrote back to his new wife Gwyneth expressing a considerable distaste for GR and the work being done in its name (reproduced in the volume {\it Surely You're Joking, Mr. Feynman}), (2) a later GRG meeting in Jena, when Zel\'dovich and Joe Weber were both elected members of the governing body, but only Joe attended its evening meeting, because at that time Soviet representatives to international organizations could only be those appointed by the government, so Jascha treated Angelica and me to sizable quantities of vodka and heavily whip-creamed desserts, and (3) the General Assembly of the International Astronomical Union in Patras, Greece in 1982, where Zel\'dovich gave one of the Invited Discourses in a reconditioned ancient Greek theatre.  This was his first time in western Europe, marking his achievement of what he called ``the second cosmic velocity.''  The first was enough for a Soviet scientist to travel in Eastern Europe (hence to the Warsaw and Jena meetings), and only the ``third cosmic velocity'' reached to the United States.

The Patras conference dinner was also the occasion on which Zel\'dovich expressed to me his puzzlement that Americans, who were allowed to read just about anything, rarely bothered, while Russians would go to a good deal of trouble to read things that were not readily available to them.  Indeed we were having a lively discussion about literature (necessarily in English!) when another conference participant came up and asked Yakov Borisovich why he wasn't doing more to help Sakharov. Suddenly his proficiency in the international language of science (broken English, he said) vanished, and he professed not to understand the question.

A later letter expressed unhappiness that we had not been present when he spoke at the US National Academy of Sciences during his first and only visit to the US.  I wrote back as soon as possible (letters between the USA and the USSR could take a long time in those days, when they were presumably read by both sides), explaining that merely being on the wrong coast in April would never have kept us away, but that neither Joe nor I was, or would ever be, a member of NAS!  I heard from Oktay H. Guseinov once after the advent of email.  He died in 2009, having spent the latter years of his career in Azerbaijan and Turkey.

The last preliminary remark was an explanation that John Archibald Wheeler might well have invented the phrase ``black hole,'' but that he could not have done so in 1967 or 1968, as is frequently written, for '``Black Holes'' in Space' is the title of a news item by Anne Ewing, published in {\it Science News Letters} in 1964 (Ewing 1964).  The item was a report of the December 1963 meeting of the American Association for the Advancement of Science, held in Cleveland.  There is also a rumor around that the phrase was heard at the very first ``Texas'' symposium in Dallas, earlier in December 1963.  Use of white holes to reverse time has never proven practical, so the phrase mush have been coined before those meetings.

\section{The Catalogue Searches}
Again one could start the story several places, but a good concrete point is the pair of papers (Zel\'dovich \& Novikov 1964, 1965), called ``Relativistic Astrophysics I and II'' and presaging much of the information later found in the two volumes of their similarly-titled book.  The papers (at least the English translations!) talk about neuron stars, continued gravitational collapse, and collapsed or frozen stars (the latter comes out as cooled stars in translation) with some calculations of likely accretion rates and energy release.  They cite the Oppenheimer papers, Baade \& Zwicky (1934) on neutron star formation as a supernova energy source, and Landau (1938) on whatever you think that paper is about.  Zel\'dovich (1964) which includes the idea of accretion on a neutron star as a source of X-ray is also relevant.

Chapter 13 of ``Relativistic Astrophysics II'' includes a footnote crediting Guseinov with the idea that detectability of compact stars is increased in binary systems.  If you think that I.S. Shklovsky should be part of this story well, so apparently did he. And I refer you to Ginzburg (1990), who outlived the other two and so got the last word, for yet another point of view.

At any rate, armed with the idea that an accreting collapsed star should be an X-ray source but contribute no significant visible light in comparison with a companion of normal luminosity, Guseinov \& Zel\'dovich (1966, submitted 18 October 1965: Zel\'dovich \& Guseinov 1966, submitted 19 November 1965) set out in search of single-line (meaning spectral features from only one star) spectroscopic binaries with invisible companions more massive than the visible stars.  They had access only to a 1948 catalogue of SBs and a smattering of later orbits, and found seven candidates.  Of the seven, four were very long period systems; none was or is a known X-ray source; and several are probably not even binary systems, but other periodic or quasi-periodic sorts of variables.  The earlier paper mentions only collapsed stars as candidates; the latter suggests a neutron star for the seventh system, Alpha Her B.

Five things happened over the next couple of years: (1) the number of known X-ray sources increased, (2) neutron stars as pulsars became part of the inventory of known objects, (3) Batten (1968) compiled a much more extensive and critical catalogue of spectroscopic binaries, (4) Kip Thorne picked up the idea of potential detectability of collapsed stars in SB1s, and (5) Trimble completed her PhD dissertation at Caltech (on motions and structures of filamentary envelope of the Crab Nebula) and had several stray weeks while the thesis was being typed, duplicated, and read by her committee before it could be defended (on 15 April 1968) and she could take off for England as a volunteer postdoc.

KS suggested that VT should go through the Batten catalogue and identify all the SB1 systems with $M_{2}$ larger than $1.4M_{\odot}$ and larger than the mass of the visible star, $M_{1}$, as estimated from its spectral type.  Barbara A. Zimmerman of the Caltech computing staff provided essential help in writing and executing the program that took the catalogued properties (period and velocity amplitude for the optical star, $P$ and $K_{1}$ plus Kepler's third law and turned them into

\begin{equation}
(M^{3}_{2} sin^{3}i)/(M_{1} + M_{2})^{2}=const \, X \, PK^{3}_{1}
\end{equation}

\noindent where $M_{2}$ is the invisible star; $M_{1}$ is the mass you have to guess from its spectral type; and $sin=1$ yields the lower limit on $M_{2}$.  It is left as an exercise for the reader to derive this (which applies only to circular orbits, the norm for short-period systems).  The presentation erroneously had $K_{2}$ in the derivation and expression (3.1) which is customarily called $f(M)$ or the mass function.

We recovered six of the seven Zel\'dovich \& Guseinov (1966) systems and added six more with $M_{2}>M_{1}$ and 38 with $M_{2}<M_{1}$, but greater than $1.4 M_{\odot}$ (the limit for stable white dwarfs).  More than 40 years downstream, none of the identified binaries is a known X-ray source (i.e. collapsed star candidate).  This was already clear at the time, because the search found about as many systems with $M_{2} > M_{1}$ (etc.) that showed eclipses, meaning that $M_{2}$ could not be a compact star.

The paper was written during a two-day stop-over I made in Chicago, where Thorne was then on sabbatical.  It is the only paper I ever hand-wrote (typing being so very much easier).  And there was a difficulty about who was to be first author.  Not I, said I, because it wasn't my idea.  Not me, said Thorne, because he hadn't done the necessary bits of arithmetic.  Should we, he then asked, call it Zel\'dovich and Zimmerman?  Well perhaps not, and Trimble \& Thorne (1969) it then became, and remained sporadically cited for a number of years, because quite a few of the stars turned out to be interesting for other reasons, though some of them weren't even really binaries and were removed from the \textit{Ninth Catalogue} (Batten 2014).

\section{What Might Have Been (the first true Z\&G system and the last wrong paper)}
Three years passed.  The Uhuru satellite was launched and, among the strong X-ray sources, Cygnus X-1 proved to be particularly, rapidly, and erratically variable (Schreier et al. 1971, Tananbaum et al. 1972).  Simultaneous flaring in X-ray and radio (Hjellming 1973) permitted optical identification with a previously-known, late O-type supergiant, HDE 226868.  Optical astronomers rushed in droves to their telescopes (Bolton 1972, Webster \& Murdin 1972; droves were smaller in those days) to check for radial velocity variability.  Sure enough, periodic, at $P=5.6$ days with a velocity half-amplitude of about $75 km/sec$, implying a minimum mass of $3.3 M_{\odot}$ for a normal O6 supergiant as the visible star. That limit has crept up over the years (Paczy\'nski 1974 and beyond). 

How could the proper response to this tale be anything but wild cheering?  Like so.  Some time between 1972 and 1999, I was on a conference tour bus seated next to Dan (Daniel Magnes) Popper.  Among the anecdotes he related was the fact that he had observed HDE 226868 as part of a radial velocity survey long ago, but had had the bad luck to observe at an integral multiple of the orbit period.  In preparing for this talk, I hunted out the relevant paper (Popper 1950).  Sure enough, there among 253 O-B6 stars with apparent magnitudes between 8.5 and 11, were color, spectral type, and radial velocity for HD 226868 (between 227704 and 226951).  The stellar velocity is tabulated as $-13 km/sec$, and of quality A, meaning that two different determinations differed by less than $4 km/sec$.  The dates of the observing runs appear, and his two spectra were taken 263 days apart, 47 times the orbit period, rather exactly.  By ill chance, any two observations could catch an unknown binary at nearly the same phase in its orbit, but I think the constancy to $4 km/sec$ out of a full aptitude of $150 km/sec$ should happen something less than 10\% of the time.  Doing a better job of this calculation is also left as an exercise for the reader.

The key point, of course, is that if Popper had prepared a good orbit for the star, it would have been in Batten's catalogue, and the 1972 application of the Zel\'dovich and Guseinov's (1966) method would have identified it as a black hole candidate.

Meanwhile, as it were, folks interested in the nature of X-ray binaries were very much aware that the mass limit for Cyg X-1 was heavily dependent on that assumed for the visible star on the basis of its spectral type.  A very massive star leaving the main sequence is not the only possibility for the combination of high temperature and low atmospheric pressure.  Post-asymptotic-giant-branch stars pass through the same conditions (though at much lower luminosity) on their way to becoming white dwarfs.  I think I may actually have been the first to think of this possibility for Cygnus X-1, for the fairly obvious reason that I had just calculated models for low mass B stars with low surface gravity (Trimble 1973) meant to describe HZ 22.  When I mentioned the idea to a stellar evolution colleague at the University of Maryland, Bill Rose, he said he thought it was a good idea and would like to be an author on the paper.  Seeing my fame as author of ``Trimble (1973)'' fade into the relative obscurity of ``Trimble \& Rose (1973)'' or even ``Rose and Trimble (1973)'' alphabetically, I co-opted my husband, Joseph Weber, as a third author of Trimble, et al. (1973).  This is the only paper I have ever had appear in print before the postcard acknowledging receipt of the manuscript arrived.  

Once again, optical observers flocked to their telescopes in droves (Margon et al. 1973, Bregman et al. 1973, droves having become slightly larger in the intervening year) to prove us wrong.  This they quickly did, getting distance estimates for the system closer to the $2 kpc$ implied by a $20 M_{\odot}$ slightly evolved O supergiant than the $200 pc$ required for a post-AGB star.  Both distances are far too large for the parallax determinations of the period, and the method used was the determination of the amount of absorption in comparison with that seen for other hot stars in the same direction of the sky.

Later work has, of course, refined knowledge of HDE 226868 = Cygnus X-1, but Trimble et al. (1973) was almost certainly the last fundamentally wrong paper on the subject.  There is a moral here for theorists:  a sufficiently definite prediction of something that can be observed will almost certainly attract the attention of colleagues wanting to prove you wrong.  Unfortunately, most of the topics discussed at IAUS 308 are too complex for such definitive predictions to be made at present.
\\
\\
Acknowledgements
\\
I am grateful to Alan Batten and Roger Griffin for e-discussions of the status of some of the Z\&G systems, to the late Dan Popper for confiding his unhappy tale long ago, to the organizers of IAUS 308 for finding time on the program for what I had to say, and to Alison Lara for her usual expert keyboarding of my usual untidy typescript.
\\
\\


\begin{thebibliography}{99}
\bibitem[Baade \& Zwicky (1934)]{BaadeZwicky34}
{Baade, W. \& Zwicky, F.} 1934,
\textit{Proc. USNAS}, 20, 254

\bibitem[Batten (1968)]{Batten68}
{Batten, A.H.} 1968,
\textit{Publ. Dom. Ap. Obs.} 8, 119

\bibitem[Batten (2014)]{Batten14}
{Batten, A.H.} 2014,
\textit{personal e-mail communication} 

\bibitem[Bolton (1972)]{Bolton72}
{Bolton, C.T.} 1972,
\textit{Nature} 235, 271; {\it Nature Phys. Sci.} 240, 124

\bibitem[Bregman (1973)]{Bregman73}
{Bregman, J. et al.} 1973,
\textit{ApJL} 186, L117

\bibitem[Dick (2013)]{Dick13}
{Dick, S.J.} 2013,
\textit{Discovery and Classification in Astronomy: Controversy and Consensus}, Cambridge Univ. Press

\bibitem[Ewing (1964)]{Ewing64}
{Ewing, A.} 1964,
\textit{Science News Letters} 8, 39 (January 18)

\bibitem[Ginzburg (1990)]{Ginzburg90}
{Ginzburg, V.L.} 1990,
\textit{ARA\&A} 20, 1

\bibitem[Guseinov \& Zeldovich (1966)]{GuseinovZeldovich66}
{Guseinov, O.H. \& Zel\'dovich, Ya.B.} 1966,
\textit{A. Zh} 43; 303 {Sov. Astr. AJ} 10, 251

\bibitem[Hjellming (1973)]{Hjellming73}
{Hjellming, R.M.} 1973,
\textit{ApJL} 182, L29

\bibitem[Landau (1932)]{Landau32}
{Landau, L.} 1932,
\textit{Phys. Z. Sowjetunion}  1, 285

\bibitem[Landau (1938)]{Landau38}
{Landau, L.} 1938,
\textit{Nature} 141, 333

\bibitem[Margon (1973)]{Margon73}
{Margon, B.} 1973,
\textit{ApJL} 185, L143

\bibitem[Oppenheimer \& Snyder (1939)]{OppenheimerSnyder39}
{Oppenheimer, J.R. \& Snyder, H.} 1939,
\textit{PR} 56 455

\bibitem[Oppenheimer \& Volkoff (1939)]{OppenheimerVolkoff39}
{Oppenheimer, J.R. \& Volkoff, G.} 1939,
\textit{PR} 55, 374

\bibitem[Paczynski (1974)]{Paczynski74}
{Paczy\'nski, B.} 1974,
\textit{A\&A} 46, 513

\bibitem[Popper (1950)]{Popper50}
{Popper, C.M.} 1950,
\textit{ApJ} 111, 495
 
\bibitem[Salpeter et al. (1971)]{Salpeter71}
{Salpeter, E.E. et al.} 1971,
\textit{ApJ} 140, 796

\bibitem[Schreier (1971)]{Schreier71}
{Schreier, E. et al.} 1971,
\textit{ApJ} 170, 121

\bibitem[Schwarzschild (1916)]{Schwarzschild16}
{Schwarzschild, K.} 1916,
\textit{Sitzber. Preus. Deut. Akad. Wiss. Berlin, KL Math.\-Phys.} 189-196 \& 424-434

\bibitem[Tananbaum et al. (1972)]{Tananbaum72}
{Tananbaum, H. et al.} 1972,
\textit{ApJL} 177, L5

\bibitem[Trimble (1973)]{Trimble73}
{Trimble, V.} 1973,
\textit{A\&A} 23, 81

\bibitem[Trimble, Rose \& Weber (1973)]{TrimbleRoseWeber73}
{Trimble, V., Rose, W.K. \& Weber, J.} 1973,
\textit{MNRAS} 162, 1p

\bibitem[Trimble \& Thorne (1969)]{TrimbleThorne69}
{Trimble, V.L. \& Thorne, K.S.} 1969,
\textit{ApJ} 156, 1013

\bibitem[Webster \& Murdin (1972)]{WebsterMurdin72}
{Webster, V.L. \& Murdin, P.} 1972,
\textit{Nature} 235, 37

\bibitem[Zeldovich (1964)]{Zeldovich64}
{Zel\'dovich, Ya.B.} 1964,
\textit{Sov. Phys. Dokl.} 9, 165

\bibitem[Zeldovich \& Guseinov(1966)]{ZeldovichGuseinov66}
{Zel\'dovich, Ya.B. \& Guseinov, O.H.} 1966,
\textit{ApJ} 144, 840; {Sov. Phys. Dokl.} 162, 791

\bibitem[Zeldovich, Ya.B. \& Novikov, I.D. (1964b)]{ZeldovichNovikov64b}
{Zel\'dovich, Ya.B. \& Novikov, I.D.} 1964b,
\textit{Sov. Phys. Dokl.} 9, 834

\bibitem[Zeldovich \& Novikov (1964)]{ZeldovichNovikov64}
{Zel\'dovich, Ya.B. \& Novikov, I.D.} 1964,
\textit{Uspekhi Phys. Nauk.} 84, 377 (= 7,763 in translation)

\bibitem[Zeldovich \& Novikov (1965)]{ZeldovichNovikov65}
{Zel\'dovich, Ya.B. \& Novikov, I.D.} 1965,
\textit{Uspekhi Phys. Nauk.} 86, 447 (= 8,522, in translation)

\bibitem[Zeldovich \& Novikov (1966)]{ZeldovichNovikov66}
{Zel\'dovich, Ya.B. \& Novikov, I.D.} 1966,
\textit{Nuovo Cimento.} I 4, 840
\end{thebibliography}
\end{document}